\documentclass[fleqn,usenatbib]{mnras}

\usepackage{newtxtext,newtxmath}

\usepackage[T1]{fontenc}

\usepackage{xcolor}
\usepackage{soul}
\DeclareRobustCommand{\VAN}[3]{#2}
\let\VANthebibliography\thebibliography
\def\thebibliography{\DeclareRobustCommand{\VAN}[3]{##3}\VANthebibliography}

\usepackage{graphicx}	
\usepackage{amsmath}	
\usepackage{xcolor}
\definecolor{revisionblue}{RGB}{0,0,0}
\definecolor{revisionbluee}{RGB}{0,90,180}

\title[]{Effect of tidal gravity and planetary rotation on the retrieved atmospheric abundances of close-in exoplanets}
\author[K. Arnav \& Hazra]{
K. Arnav,$^{1}$
Gopal Hazra,$^{1}$\thanks{E-mail: hazra@iitk.ac.in}
\\
$^{1}$Department of Physics, Indian Institute of Technology Kanpur, Kanpur 208016, India.
}

\date{Accepted XXX. Received YYY; in original form ZZZ}

\pubyear{\the\year{}}

\begin{document}
\label{firstpage}
\pagerange{\pageref{firstpage}--\pageref{lastpage}}
\maketitle

\begin{abstract}
Most modern atmospheric retrievals adopt the simplifying assumption that the planetary atmosphere endures no planetary rotation and stellar tidal effect. However, for tidally locked close-in exoplanets, the gravitational influence of the host star and the rapid rotation of the planet can significantly modify the effective gravity, leading to changes in the atmospheric scale height and mixing ratios of molecular abundances. In this work, we develop a combined framework to include these rotation and tidal effects into a forward and retrieval model to study how they affect the molecular abundances of close-in exoplanets. We specifically apply our model to the planet WASP-12b, observed with HST, and WASP-121b, observed with JWST, and investigate how atmospheric retrieval parameters change when tidal and centrifugal corrections to gravity are included. The forward model calculation for strongly affected gravity in WASP-12b shows an increment in transit depth in the range of 300-600 ppm for major molecules, whereas for WASP-121b with relatively less gravity reduction shows variations of {60-140} ppm. The atmospheric retrievals for both planets with and without effective gravity corrections show differences in the retrieved molecular abundances. {To validate the detectability of these gravity-induced changes, we perform synthetic injection-recovery experiments over a broad wavelength range, which show statistically significant shifts in the retrieved atmospheric parameters for simple atmospheric compositions, while more complex atmospheric compositions exhibit smaller changes due to increased retrieval degeneracies. Finally, we investigate the robustness of these results using non-isothermal P-T profiles and cloudy atmospheric models, and find that the inferred trends remain qualitatively consistent.}

\end{abstract}



\section{Introduction}
Transmission spectroscopy has emerged as a primary tool for probing the chemical compositions and thermal structures of exoplanetary atmospheres, particularly for hot Jupiters whose large radii and high temperatures produce prominent spectral signatures. Earlier, the Hubble Space Telescope (HST) was instrumental in detecting and characterizing various molecules in exoplanet atmospheres, primarily using transmission spectroscopy (0.2–1.7 $\mu m$) to analyze light passing through the atmosphere during transit, with key detections including water ($H_2O$), methane ($CH_4$), carbon dioxide ($CO_2$) and sodium ($Na$) \citep[e.g.,][]{Swain_2014, Murphy2025} along with studies of high-temperature clouds, haze \citep{Gibson2012, Kreidberg2014, Kreidberg2018} and atmospheric escape \citep{Spake2018, Hazra2025}. The transition from the Hubble Space Telescope (HST) to the James Webb Space Telescope (JWST),  with its 0.6 to 28 $\mu m$ coverage and higher spectral resolution, allows for the definitive identification of atmospheric species by capturing multiple molecular bands simultaneously and demonstrated the ability to achieve a spectroscopic noise floor of $< 50$ parts per million (ppm), performing very close to theoretical photon-limited noise in a single transit \citep[e.g.,][]{Rustamkulov2023}. Traditionally, atmospheric models used in transmission spectroscopy commonly assume spherical symmetry and a constant surface gravity. However, as the field transitions from the era of the Hubble Space Telescope (HST) to the unprecedented precision and spectral coverage of the James Webb Space Telescope (JWST), we can investigate how these simplifying assumptions, which do not hold true, affect atmospheric parameters. Recently \citet{Banerjee_2023} has highlighted that for close-in, fast-rotating planets like WASP-76b and WASP-121b, centrifugal forces arising from rapid rotation can lead to departures from spherical symmetry and introduce biases in retrieved atmospheric parameters. 

In addition to centrifugal acceleration, another physical effect commonly neglected in atmospheric retrievals is the modification of the effective gravitational acceleration by tidal interactions with the host star. There are many planets for which the tidal deformation due to the host star has been observed \citep{Burton2014, Barros_et_al2022, Akinsanmi2019, Akinsanmi2024b}.  WASP-12b is one such planet that has shown a strong tidal distortion \citep{Li_2010, Akinsanmi2024b}. For close-in, tidally locked exoplanets, centrifugal acceleration from rotation and tidal acceleration induced by the stellar gravitational field both reduce the effective gravity experienced by the atmospheric limb. The resulting change in effective gravity alters the height of the atmospheric scale, thereby directly influencing the amplitude of spectral features in the transmission spectra. In modern atmospheric retrieval codes (see \citet{MacDonald2023} for a description of all available retrieval codes), such corrections are generally not incorporated into exoplanet atmospheric retrieval frameworks. This raises a fundamental question: Does the inclusion of centrifugal and tidal corrections to the effective gravity lead to measurable biases in retrieved atmospheric parameters when interpreting high-precision transmission spectra?

In this work, we address this question by systematically quantifying the impact of centrifugal and tidal gravity corrections on transmission spectra and atmospheric retrievals of hot Jupiters. As the tidal force is dominant for short-period planets and the centrifugal force is stronger for fast-rotating planets, we chose short-period planets with high rotation rates so that the effects of both tides and rotation become significant. We focus on the close-in ultra-hot Jupiters WASP-12b, a highly irradiated planet experiencing strong tidal and rotational effects, and {WASP-121b}, for which high-quality JWST transmission spectroscopy data are currently available. Using the atmospheric retrieval code \textsc{POSEIDON}, {we first investigate the influence of reduced effective gravity on synthetic transmission spectra and then perform controlled synthetic injection-recovery experiments to assess how gravity-induced changes propagate into the retrieved atmospheric parameters. We subsequently perform atmospheric retrievals with and without gravity corrections using HST observations of WASP-12b and JWST observations of WASP-121b under multiple atmospheric assumptions, including non-isothermal P-T profiles and cloudy models, to examine the robustness of the inferred atmospheric abundances. This study aims to assess when gravity corrections become significant, how they interact with atmospheric model degeneracies, and what their inclusion implies for the interpretation of current and future exoplanet transmission spectra.}

The paper is organized as follows. In the next section, we present the analytical formulation for incorporating centrifugal acceleration and tidal gravity into the effective gravitational acceleration and describe the forward-model and atmospheric retrieval framework adopted throughout this work. {In Section~3, we first investigate the impact of gravity corrections on synthetic transmission spectra, followed by controlled synthetic injection-recovery experiments that validate the retrieval methodology. We then apply the framework to HST and JWST observations of WASP-12b and WASP-121b, respectively. We also perform an experiment with enhanced gravity reduction with WASP-121b JWST spectra in addition to the real gravity reduction observed in the planet to quantify maximum effect of reduced gravity on retrieved molecular abundances. The robustness of the inferred atmospheric parameters under different retrieval assumptions, including non-isothermal pressure-temperature profiles, and cloudy atmospheres are also presented. Finally, Section~4 summarizes the principal conclusions and discusses the implications of our results for future atmospheric retrieval studies of close-in exoplanets.}

\section{Incorporation of the combined effect of centrifugal acceleration and tidal gravity}
In transmission spectroscopy, exoplanets are generally assumed to be spherical, which is a reasonable approximation. However, for some close-in planets such as WASP-76b, WASP-121b, and WASP-19b, which are fast rotators, this assumption can lead to inaccuracies in predicted transmission spectra and the atmospheric parameters retrieved \citep{Banerjee_2023}. The fact that rapid rotation significantly affects the planetary shape is evident from the substantial difference between the polar and equatorial radii of such planets \citep{guillot2009giantplanets, Houdayer2023}. While the effect of rotation on the effective gravitational acceleration is well known and routinely accounted for in models of fast-rotating Solar System giant planets \citep[e.g.,][]{Correia2003}, it has not been incorporated into atmospheric retrieval models for exoplanets. In addition, the stellar tide, which plays an important role in atmospheric circulation and planetary deformation \citep{Akinsanmi2019, Akinsanmi2024b}, is not accounted for when retrieving atmospheric abundances from transmission spectra.

\begin{figure}
	\includegraphics[width=1.2\columnwidth]{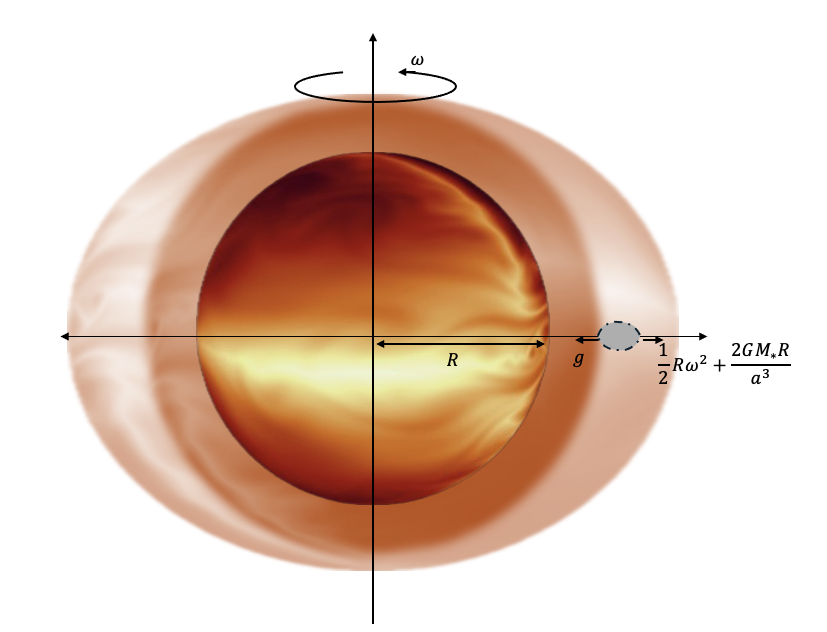}
    \caption{A schematic cross section of a planet along the plane of the terminator showing the accelerations acting on an atmospheric parcel (light grey). The atmospheric extent corresponding to the scenario with rotation and tidal effects included is illustrated by a light orange elliptical annulus (outer). The atmospheric scale heights in the diagram have been exaggerated for visibility. The atmospheric extent corresponding to the scenario without the rotation and tidal effects included is illustrated by a darker orange circular annulus (inner). The atmospheric parcel is subjected to both centrifugal and tidal accelerations. The centrifugal acceleration acts perpendicular to the axis of rotation, while the tidal acceleration arises from the differential gravitational pull of the host star. The net radial component of these accelerations opposes the planetary gravitational acceleration $g$. In the quantitative analysis presented in this paper, the atmospheric annulus relevant to transmission spectroscopy is approximated as circular.}
    \label{fig:centrifugal}
\end{figure}

\subsection{Effect of Centrifugal acceleration}
\label{sec:rotation}
In a schematic diagram (as shown in Figure~\ref{fig:centrifugal}), we depict how centrifugal force and tidal force could act on an atmosphere of the planet. We show a schematic cross-section of the planet along the plane of the terminator, illustrating the accelerations acting on an atmospheric parcel for both centrifugal and tidal forces. The planet is assumed to be tidally locked, such that the orbital period is equal to the rotation period, and the planetary rotation axis is taken to be perpendicular to the orbital plane. Under these assumptions, one can compute the latitude-dependent effective gravitational acceleration, accounting for the centrifugal contribution \citep{Banerjee_2023}. The average net gravitational acceleration, accounting for centrifugal force relevant to transmission spectroscopy, was obtained by integrating over latitudes from the equator to the pole and dividing by the corresponding latitude range. Assuming hydrostatic equilibrium in the atmosphere, the resulting effective gravitational acceleration due to centrifugal effect is given by
\begin{equation}
    g_c'(R) = g(R) - \frac{1}{2}\omega^2 R
    \label{eq:centrifugal}
\end{equation}
where $g_c'(R)$ is the corrected effective gravity, g(R) is the standard gravity assuming the sphericity of the planet, $\omega$ is the angular velocity of the planet, and $R$ is the radius of the planet.




\subsection{Tidal Effects}\label{sec:tidal}

For tidally locked planets, the host star can significantly distort the
planetary gravitational field and influence atmospheric dynamics, with
potential implications for the atmospheric structure and chemical
abundances. The dynamics of tidal potentials and dissipation are discussed
in detail by \citet{Ogilvie_2004}, whose formalism we follow to
obtain a closed-form expression for the tidal potential. The stellar
gravitational potential, expanded about the planet's center up to
quadrupole order, is given by

\begin{equation}
\Phi_{\rm tide}
=
\frac{GM_\star}{2a^3}
\left[
r^2-3\left(\mathbf{r}\cdot\hat{\mathbf{a}}\right)^2
\right],
\label{eq:tidal_potential}
\end{equation}

where $G$ is the universal gravitational constant, $M_\star$ is the stellar
mass, $a$ is the star-planet separation {(taken to be the orbital
semi-major axis in our analysis), $\mathbf{r}$ is the position vector of an
atmospheric parcel relative to the planet's center, and
$\hat{\mathbf{a}}$ is the unit vector pointing from the planet towards the
star.}

\begin{table*}
\centering
\caption{List of target planets with strong tidal forces and fast rotation. 
The selected planets are favorable candidates for gravity-correction effects owing to their low surface gravity, large planetary radii, short orbital periods, small semi-major axes (enhancing tidal effects), and the availability of transmission spectroscopy observations.}
\label{tab:target}
\begin{tabular}{lcccccccc}
\hline
Exoplanet &
$R/R_{\rm J}$ &
$M_\star/M_\odot$ &
$T_{\rm eq}$ (K) &
$a$ (AU) &
Period (days) &
$g$ (m s$^{-2}$) &
$g'$ (m s$^{-2}$) & Changes ($\%$) \\
\hline
WASP-76b   & 1.854 & 1.458 & 2226.25 & 0.033 & 1.809 & 6.45 & 5.92 & 8.1 \\
WASP-39b   & 1.279 & 0.193 & 1166.43 & 0.048 & 4.055 & 3.91 & 3.84 & 1.8 \\
HAT-P-32b  & 1.98  & 1.132 & 1835.66 & 0.034 & 2.150 & 4.30 & 3.84 & 10.7\\
WASP-107b  & 0.94  & 0.683 & 744.74  & 0.055 & 5.721 & 2.69 & 2.67 & 0.7 \\
HD~209458b & 1.39  & 1.23  & 1476.81 & 0.047 & 3.524 & 9.37 & 9.25 & 1.3\\
WASP-121b  & 1.742 & 1.33 & 2409.29 & 0.026 & 1.274 & {9.56} & {8.62} & {9.9}\\
WASP-12b   & 1.965 & 1.325 & 2567.38 & 0.023 & 1.091 & 9.44 & 7.95 & 15.8\\
\hline
\end{tabular}
\end{table*}

\begin{figure}
    \centering
    \includegraphics[width=1.0\linewidth]{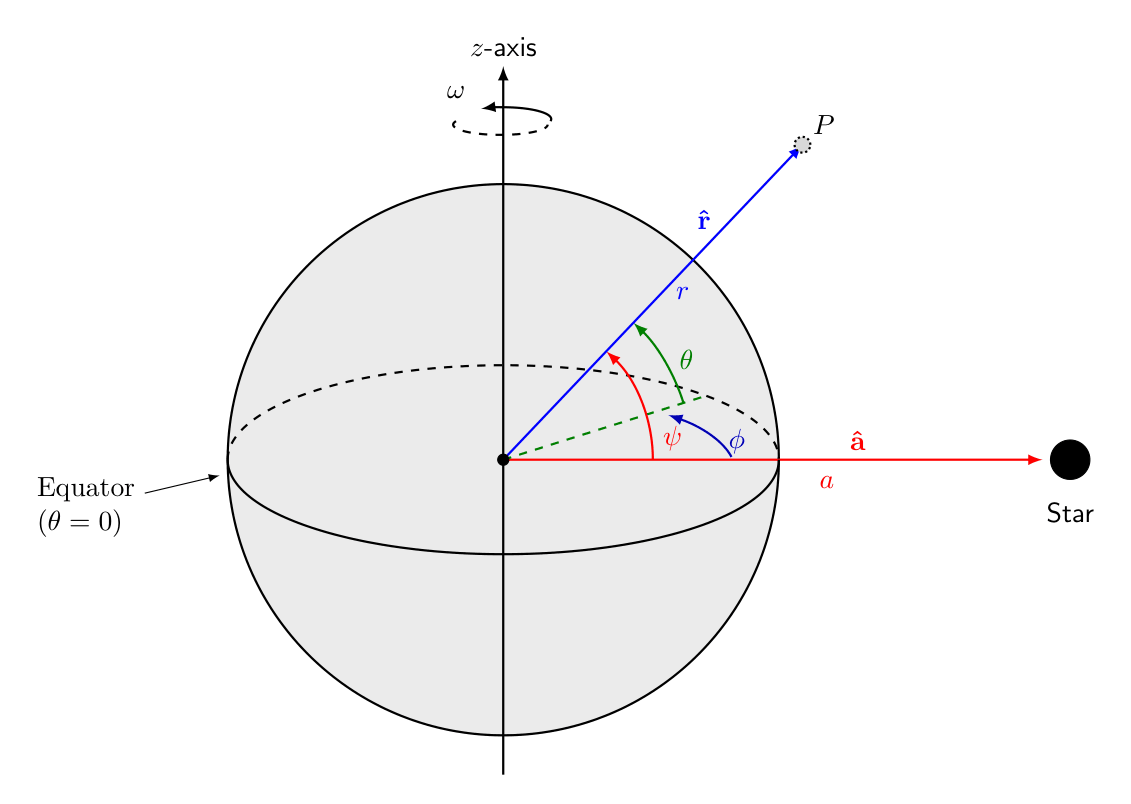}
    \caption{{Schematic geometry used to define the coordinate system and
    the angles entering the tidal correction. The planetary rotation axis
    is shown by the $z$-axis, while $\hat{\mathbf a}$ denotes the
    star--planet direction and $\mathbf r$ the position vector of an
    atmospheric parcel $P$. The latitude $\theta$ is measured from the
    planetary equatorial plane, the longitude $\phi$ is measured from star-facing direction and $\psi$ is the angle between the local
    radial direction $\hat{\mathbf r}$ and the star--planet axis. The
    equatorial projection of $\hat{\mathbf r}$ is shown by the green
    dashed line.}}
    \label{fig:tidal_geometry}
\end{figure}

{To clarify the geometrical convention used throughout this work, we define
$\psi$ as the angle between the local radial direction
$\hat{\mathbf{r}}=\mathbf{r}/r$ and the star--planet axis
$\hat{\mathbf{a}}$, as illustrated in Figure~\ref{fig:tidal_geometry}.
Thus,}

{\begin{equation}
\mathbf{r}\cdot\hat{\mathbf{a}}
=
r\cos\psi.
\label{eq:rdotahat}
\end{equation}
}
{We describe the position of the atmospheric parcel using spherical
coordinates $(r,\theta,\phi)$, with an important convention that
$\theta$ denotes latitude measured from the planetary equatorial
plane, following the convention adopted by \citet{Banerjee_2023}. The
longitude $\phi$ is measured in the equatorial plane from the
star--planet axis. For the zero-obliquity configuration considered here,
the planetary rotation axis is perpendicular to the star--planet axis,
so that the latter lies in the planetary equatorial plane. We may therefore
choose the instantaneous $x$-axis along $\hat{\mathbf{a}}$, giving}

{\begin{equation}
\hat{\mathbf{r}}
=
\left(
\cos\theta\cos\phi,\,
\cos\theta\sin\phi,\,
\sin\theta
\right),
\label{eq:rhat_latitude}
\end{equation}
}
{and}

{\begin{equation}
\hat{\mathbf{a}}=(1,0,0).
\label{eq:ahat}
\end{equation}
}
{Taking the scalar product of these two unit vectors gives}

{\begin{equation}
\cos\psi
=
\hat{\mathbf{r}}\cdot\hat{\mathbf{a}}
=
\cos\theta\cos\phi.
\label{eq:cospsi}
\end{equation}
}
{More generally, if $\Phi$ denotes the longitude of the star--planet axis
in an arbitrary longitude convention, this relation can be written as
$\cos\psi=\cos\theta\cos(\Phi-\phi)$. Our choice of longitude origin along
the instantaneous star--planet axis simply sets $\Phi=0$.}

{Using Eq.~\ref{eq:rdotahat}, the tidal potential can therefore be written
in terms of the latitude and longitude of the atmospheric parcel as}

{\begin{equation}
\Phi_{\rm tide}
=
\frac{GM_\star r^2}{2a^3}
\left[
1-3\cos^2\theta\cos^2\phi
\right].
\label{eq:tidal_potential_spherical}
\end{equation}
}
{The corresponding radial tidal acceleration is obtained from
$a_{\rm tide,r}=-\partial\Phi_{\rm tide}/\partial r$, yielding}

{\begin{equation}
a_{\rm tide,r}(\theta,\phi)
=
\frac{GM_\star r}{a^3}
\left(
3\cos^2\theta\cos^2\phi-1
\right).
\label{eq:tidal_acceleration_general}
\end{equation}
}
{For the atmospheric annulus considered in our transmission-spectrum
model, we seek a characteristic correction that captures the maximum
outward radial contribution of the tidal field. The quadrupolar tidal
acceleration reaches its maximum when $\cos^2\psi=1$, i.e. when
$\psi=0$, corresponding to the substellar point. Under the zero-obliquity
configuration adopted above, the substellar point lies on the planetary
equator, and hence}

{\begin{equation}
\theta=0,\qquad\phi=0,\qquad\psi=0.
\label{eq:max_tidal_geometry}
\end{equation}
}
{Thus, for the specific configuration used to estimate the maximum tidal
effect,}

{\begin{equation}
a_{\rm tide}
=
\left.a_{\rm tide,r}\right|_{\theta=0,\phi=0}
=
\frac{2GM_\star R}{a^3},
\label{eq:tidal_acceleration_max}
\end{equation}
}
{where $R$ is the planetary radius at the atmospheric level relevant to
the calculation. We emphasize that $\theta$ and $\psi$ are distinct
angles in general; their simultaneous value of zero arises only for the
specific substellar, zero-obliquity configuration adopted here. Other
locations on the planet experience a different, generally smaller
outward tidal contribution, and the full spatially varying tidal field is
not represented by this one-dimensional characteristic correction. We
adopt the maximum outward value because our aim is to investigate the
largest plausible impact of tidal acceleration on the atmospheric
retrieval parameters.} The steep dependence of orbital distance on tidal stress makes the close-in exoplanets more prone to tidal distortion, even for low-mass host stars. Effectively, the tidal stress from the host star reduces the gravitational field of the planetary body and redistributes the pressure of the hydrostatic atmosphere.

{The centrifugal contribution is treated separately by averaging its
radial component over latitude. For a synchronously rotating planet,
the latitude-dependent radial centrifugal acceleration is
$a_{\rm cent,r}=\omega^2R\cos^2\theta$, whose latitude average is
$\langle a_{\rm cent,r}\rangle=\frac{1}{2}\omega^2R$. Combining this
latitude-averaged centrifugal contribution with the characteristic
maximum tidal contribution gives the modified effective gravitational
acceleration,}

\begin{equation}
g'
=
g-\frac{1}{2}R\omega^2
-\frac{2GM_\star R}{a^3}.
\label{eq:modified_gravity}
\end{equation}

where R = radial distance from the planet center, $\omega$  = rotation rate of the planet, and g is the acceleration due to gravity for the planet mass alone. The atmospheric pressure scale height $H = \frac{kT}{\mu g'}$ is inversely proportional to effective gravitational acceleration and thus increases when both the effect of centrifugal force and tidal force are included in the effective gravity calculation. The modified atmospheric scale height, in turn, affects the pressure distribution in the atmosphere and restructures the abundances of atmospheric species and hence changes in the observed transmission spectra from the exoplanets. A list of target planets that are prone to tidal and rotational effects is provided in Table~\ref{tab:target}. 


In order to quantify the physical impact of gravity modifications on atmospheric structure, we compute the gravitational energy stored in a hydrostatic planetary atmosphere. The atmosphere is assumed to be in vertical hydrostatic balance,
\begin{equation}
    \frac{dP}{dz} = -\rho g_{eff}
    \label{eq:hydrostatic}
\end{equation}
where $P$ is pressure, $\rho$ is mass density, $z$ is altitude above a reference radius $R_p$, and $g_{eff}$ is the effective gravitational acceleration. We also assume the atmosphere to be isothermal at temperature $T$. The pressure, density and temperature are related by the ideal gas equation of state,
\begin{equation}
    P = \rho\frac{ k T}{\mu m_H}
\end{equation}
where $k$ is the Boltzmann constant, $\mu$ is the mean molecular weight (in proton masses) and $m_H$ is the hydrogen mass. The gravitational potential energy stored in the atmosphere per unit surface area is given by,
\begin{equation}
E_{\mathrm{grav}} = \int_{0}^{\infty} \rho(z)\, g_{\mathrm{eff}}\, z \, \mathrm{d}z .
\end{equation}
Here, we assume zero as the reference point on the surface of the planet. Combining the hydrostatic relation with the ideal gas equation of state, the gravitational energy stored per unit surface area is given by,
\begin{equation}
    E_{\mathrm{grav}} = \frac{P_{\mathrm{ref}} k T}{\mu m_H g_{\mathrm{eff}}}
\end{equation}
where $P_{\mathrm{ref}}$ is the reference pressure at the reference point, which is taken to be $10^6 Pa$ for further retrievals. Taking $\mu = 2.3$, $m_H = 1.673 \times 10^{-27} kg$, we calculated the heat budget for WASP-12b. The planetary properties and the reduced gravity values are given in Table~ \ref{tab:target}. The calculated gravitational potential energy stored per unit surface area for WASP-12b, using the standard surface gravity
$g = 9.4362\,\mathrm{m\,s^{-2}}$, is $9.56 \times 10^{11}\,\mathrm{J\,m^{-2}}$, whereas for the gravity-corrected case with
$g' = 7.9447\,\mathrm{m\,s^{-2}}$ the energy increases to
$1.14 \times 10^{12}\,\mathrm{J\,m^{-2}}$. We assume equilibrium temperatures and a hydrogen-dominated atmosphere while estimating the gravitational energy stored in the atmospheric column for both planets. For WASP-12b, the inclusion of centrifugal and tidal corrections increases the atmospheric energy budget by nearly 20\%. {Similarly, for WASP-121b, using $g=9.5565~\mathrm{m\,s^{-2}}$ and the gravity-corrected value $g'=8.6275~\mathrm{m\,s^{-2}}$, we find that the gravitational energy stored per unit surface area increases from $9.05\times10^{11}$ to $1.00\times10^{12}~\mathrm{J\,m^{-2}}$, corresponding to an increase of approximately $11\%$. Thus, the inclusion of centrifugal and tidal corrections results in an increment in the atmospheric gravitational energy budget for WASP-121b, consistent with the reduction in its effective gravity.}



\subsection{Retrieval setup including tidal gravity and centrifugal term}
\label{subsec:retrieval_setup}
For our calculations, we use the \textsc{POSEIDON} atmospheric retrieval framework to generate synthetic transmission spectra and to perform atmospheric retrievals. \textsc{POSEIDON} \citep{MacDonald2023, MacDonald2022} is a Python-based atmospheric retrieval framework designed to infer exoplanet atmospheric properties from transmission and emission spectroscopy. {In this work, all forward models and retrievals are performed in the opacity sampling mode using pre-computed molecular cross sections, which provides an efficient treatment of molecular and continuum opacities for low- and medium-resolution transmission spectra.} The molecular opacity database is primarily drawn from ExoMol \citep{Tennyson_2016}, together with the corresponding continuum opacity sources implemented within \textsc{POSEIDON}. Atmospheric retrievals are carried out using a nested sampling approach, which enables efficient exploration of potentially non-Gaussian, multimodal posterior distributions. We employ the \texttt{pymultinest} package \citep{Buchner_2014}, based on the \textsc{MultiNest} algorithm \citep{2004AIPC..735..395S, Feroz_2009}. 

Our main aim is to determine whether the reduction in effective gravity due to centrifugal and tidal forces can alter the observed transit spectra by reorganizing pressure and atmospheric abundances. As a first step, we use the forward model to generate synthetic spectra for two planets, WASP-12b and {WASP-121b}, with reduced effective gravity, as shown in the last two rows of Table~\ref{tab:target}. We expect to see changes in the synthetic spectra as the integrated atmospheric column stores more gravitational energy when the effective gravity is reduced (as explained in sec~\ref{sec:tidal}), owing to the redistribution of atmospheric mass over a larger number of scale heights.


For all synthetic models and retrievals presented in this work, unless mentioned, we assume a cloud-free, isothermal atmosphere in hydrostatic equilibrium, composed of a hydrogen-dominated background with a fixed He:H$_2$ ratio of 0.17. The atmosphere is assumed to be laterally homogeneous, with uniform temperature and composition at all longitudes and latitudes, varying only in the radial direction. Ray deflection due to refraction or scattering is neglected for all impact parameters and azimuthal angles.
For WASP-12b, synthetic models are constructed using planetary properties from Exomast \citep{Exomast}, with prior expectations on molecular abundances guided by \citet{Himes2022}. The main spectral features used for the synthetic spectra are {[`CO', `CO2', `CH4', `H2O', `NH3', `TiO']} with the bulk species being [`H2', `He']. {For WASP-121b, synthetic spectra are generated using planetary parameters and prior information from \citet{Saha_2025}. The main spectral features used are [`H2O', `CO', `SiO', `VO', `TiO'].} These synthetic spectra are intended solely to examine relative changes between models with and without gravity corrections, and are not designed to reproduce the observed transmission spectra in detail.

While synthetic spectra would help us directly quantify the effects of stellar tidal gravity and rotation-induced centrifugal force on transit spectra, they cannot tell us about atmospheric abundances. We need retrievals to estimate atmospheric abundances from observed transit spectra, incorporating tidal and rotational corrections. For our study, retrievals are performed over a pressure range of $10^{-6}$-100~bar, discretized into 100 uniformly spaced layers in log-pressure. The planetary reference radius is defined at a pressure of 10~bar. A constant-resolution wavelength grid with resolving power {$R=\lambda/\Delta\lambda=15000$} is used for the forward model, with spectra subsequently binned to the instrumental resolution during likelihood evaluation. 

{Following the forward-model calculations, we perform a series of synthetic injection-recovery experiments to validate the influence of the gravity correction on the inferred atmospheric parameters. Such controlled experiments were previously employed by \citet{Banerjee_2023} to demonstrate that atmospheric retrievals are able to recover the injected atmospheric properties while quantifying the impact of centrifugal acceleration on the retrieved abundances. In the present work, we extend this methodology by simultaneously incorporating both centrifugal and tidal contributions to the effective gravity, allowing us to isolate and quantify their combined influence on atmospheric retrievals before analysing real observations.}

{For these experiments, synthetic transmission spectra are first generated for analogue atmospheres representative of WASP-12b and WASP-121b using prescribed atmospheric properties as the injected (true) parameters, which included the modification in g. Atmospheric retrievals are subsequently performed on these simulated datasets using both the standard and gravity-corrected formulations in order to quantify the resulting changes in the recovered atmospheric parameters, which we present in the next section}


\section{Effect of centrifugal acceleration and tidal gravity on atmosphere}\label{sec:results}
\subsection{Synthetic Spectra of WASP-12b and {WASP-121b}}
\begin{figure*}
\centering

\begin{minipage}{0.42\textwidth}
\centering
\textbf{(a) WASP-12b}
\vspace{0.1cm}
\includegraphics[width=\linewidth]{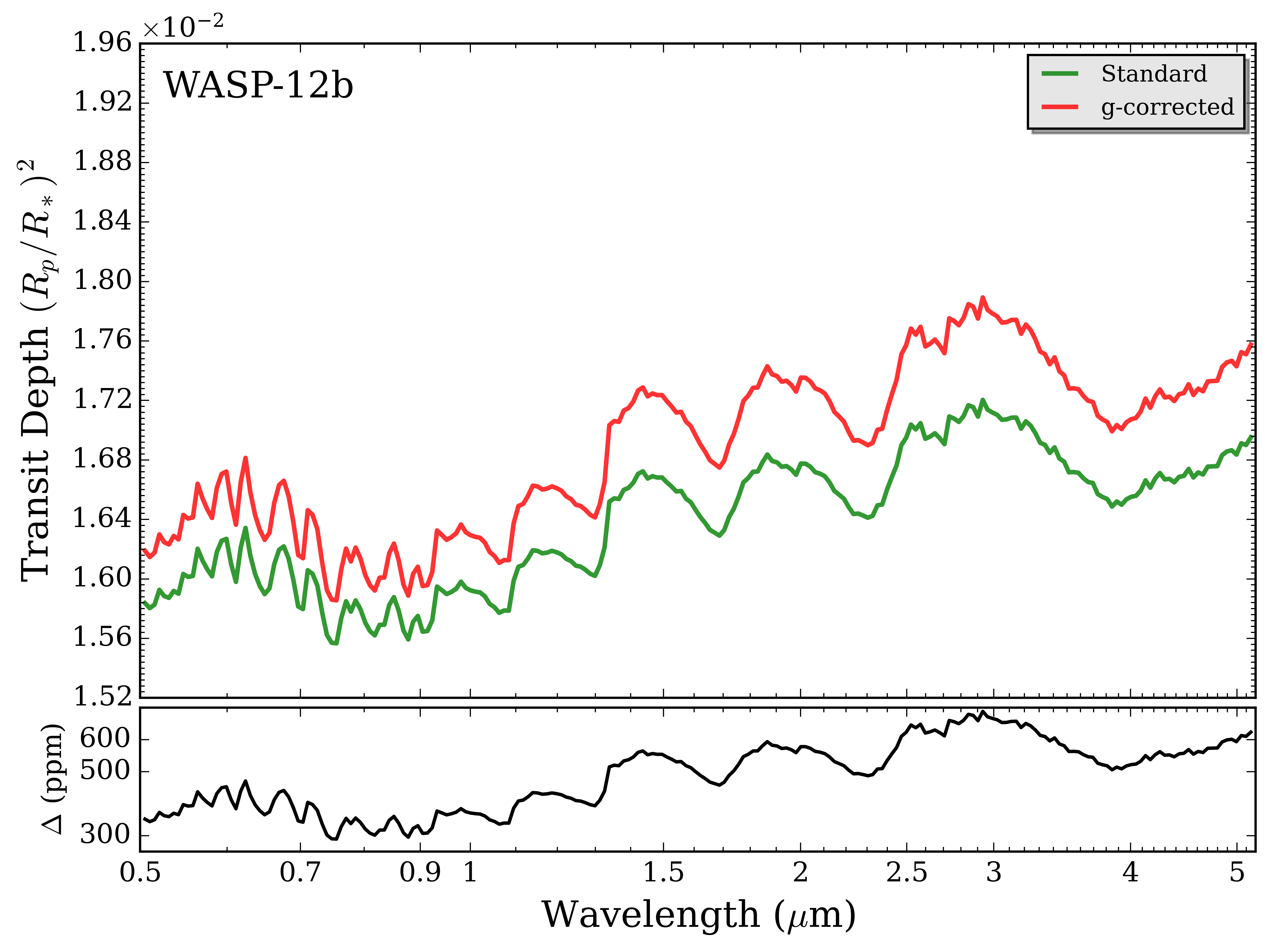}
\end{minipage}
\begin{minipage}{0.42\textwidth}
\centering
\textbf{(b) WASP-121b}
\vspace{0.1cm}
\includegraphics[width=\linewidth]{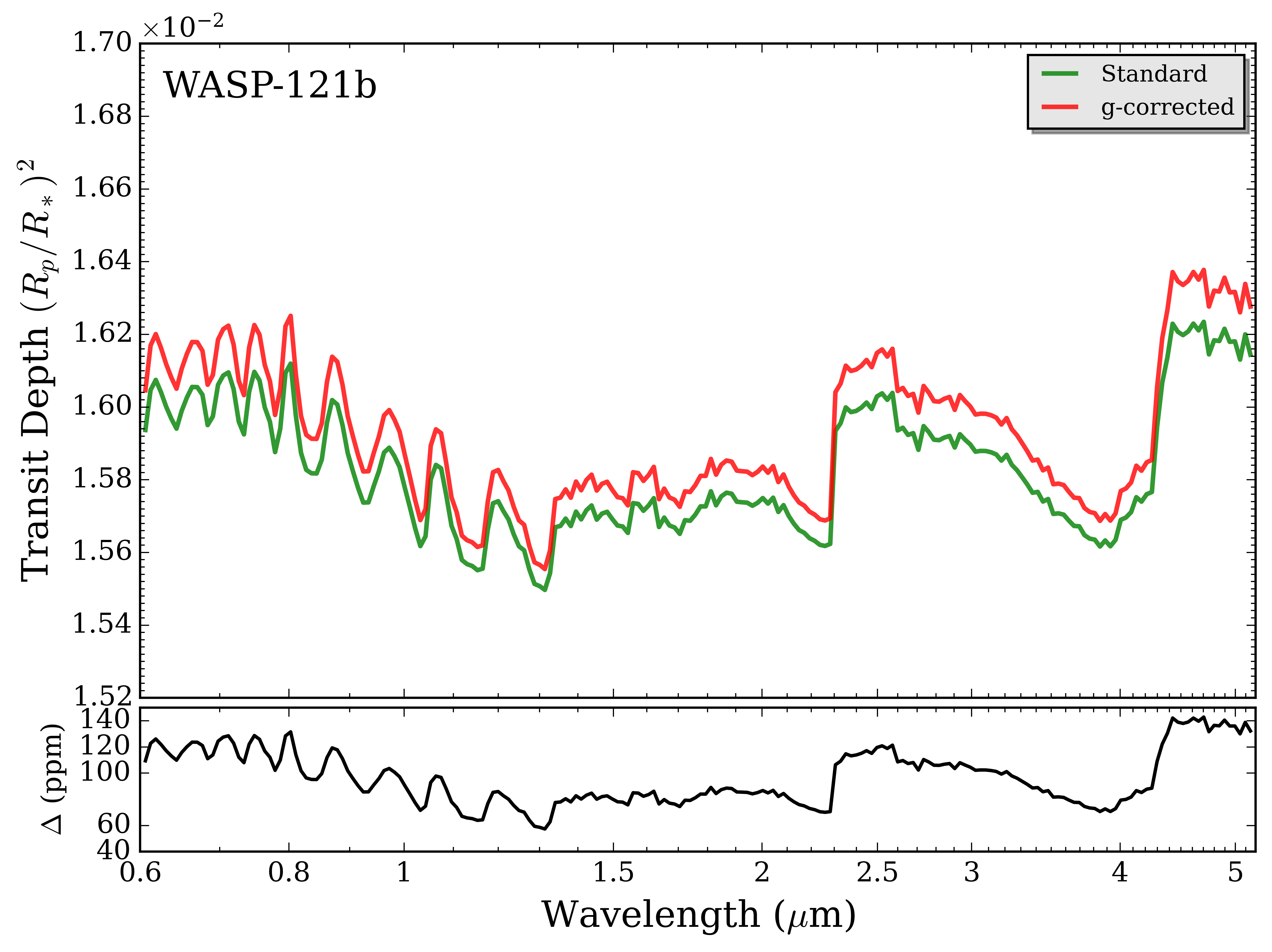}
\end{minipage}

\caption{
Forward-model transmission spectra illustrating the impact of centrifugal and tidal gravity corrections. Left and right Panels correspond to WASP-12b and {WASP-121b}, respectively. In each panel, the upper panel shows the transmission spectra computed with and without gravity corrections in the {red and green solid lines}, respectively, while the lower panel shows the corresponding difference in transit depth ($\Delta$ppm). The inclusion of gravity corrections leads to an increase in the effective atmospheric scale height for different molecules, producing a wavelength-dependent enhancement in the transit depth.
}
\label{fig:gravity_forward_models}
\end{figure*}

Here we present resulting synthetic transit spectra from the forward model for WASP-12b and {WASP-121b} to compare the resulting transmission spectra for the standard gravity case and the case with both centrifugal and tidal corrections. The forward-model calculations indicate that, in the presence of strong tidal forces and significant centrifugal acceleration due to rapid rotation, a planet can exhibit enhanced transit depths in transmission spectra as a consequence of effective molecular abundances spread over the atmospheric thickness of the planet. Figure~\ref{fig:gravity_forward_models} shows the synthetic transmission spectra of WASP-12b and {WASP-121b} with and without the inclusion of tidal and rotational gravity corrections. The left and right panels correspond to WASP-12b and WASP-121b, respectively, with spectra computed using the standard gravity shown in {green} and those including gravity corrections shown in {red}. Owing to the stronger reduction in effective gravity for WASP-12b, we observe significantly larger gravity-induced changes in the transmission spectrum, at the level of {$\sim$300-600~ppm}, particularly at prominent molecular absorption features. In contrast, the corresponding changes for {WASP-121b} are smaller, at the level of {$\sim$60-140~ppm.}

The increase in transit depth with reduced gravity is expected, as the atmospheric scale height $H$ is inversely proportional to the gravitational acceleration $g$. Also, we notice that the transit depth changes with the wavelength as well. This is due to the different responses of molecules with different mean molecular weights to the reduced gravitational acceleration, which redistributed the atmospheric molecules differently over a larger scale height.

{The forward-model spectra presented above demonstrate that centrifugal and tidal gravity corrections produce measurable, wavelength-dependent changes in the transmission spectra of both WASP-12b and WASP-121b. We now investigate whether these spectral differences translate into measurable shifts in the atmospheric parameters inferred from atmospheric retrievals. To do this, we perform synthetic injection-recovery experiments for analogue atmospheres of both planets, comparing retrievals performed with the standard and gravity-corrected formulations.}

{\subsection{Synthetic injection-recovery tests}}
{\subsubsection{WASP-12b}}
{For the WASP-12b analogue, we assume a cloud-free, isothermal atmosphere with a temperature of $T = 2500$~K and active opacity sources due to H$_2$O and CO, with injected volume mixing ratios of $\log(X_{\mathrm{H_2O}}) = -2$ and $\log(X_{\mathrm{CO}}) = -4$. The synthetic transmission spectrum is generated over the wavelength range 0.5--5.0~$\mu$m and sampled into 228 spectral points, each assigned a uniform uncertainty of 30~ppm. This simple atmospheric composition allows us to investigate the influence of the gravity correction in a minimally degenerate retrieval framework.}

\begin{figure*}
	\includegraphics[width=1.0\textwidth]{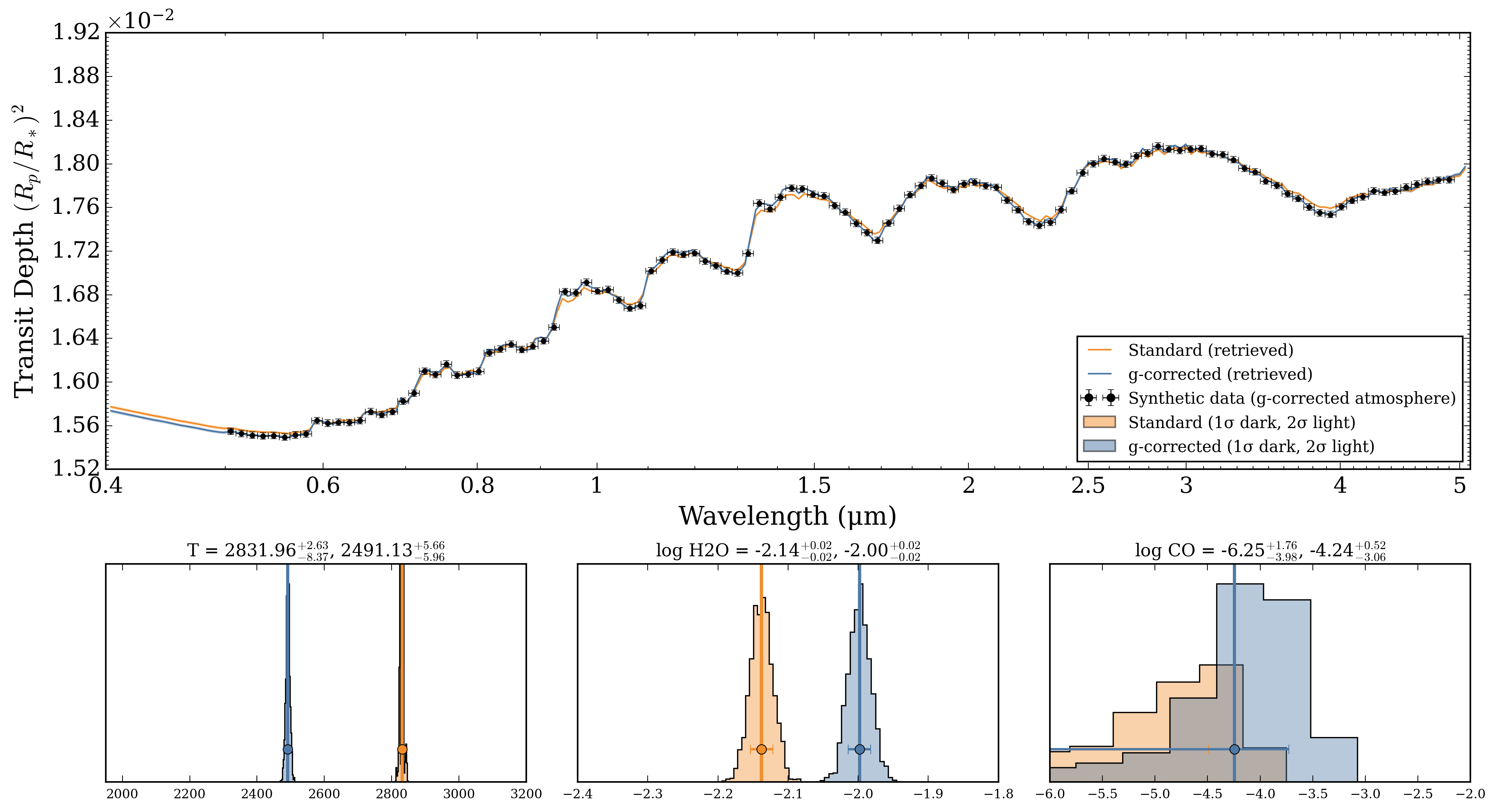}
    \caption{{Synthetic injection-recovery test for a WASP-12b analogue using an isothermal, cloud-free atmosphere. The black points with horizontal and vertical error bars show the synthetic transmission spectrum generated from the gravity-corrected atmosphere, with the synthetic observations constructed at the adopted spectral resolution and noise level. The upper panel compares the median retrieved transmission spectra obtained using the standard-gravity formulation ($g = 9.4362~\mathrm{m\,s^{-2}}$; orange) and the gravity-corrected formulation ($g' = 7.9447~\mathrm{m\,s^{-2}}$; blue). The darker and lighter shaded regions surrounding each retrieved spectrum denote the $1\sigma$ and $2\sigma$ posterior credible intervals, respectively. The lower panels show the corresponding marginal posterior distributions for the retrieved atmospheric temperature, $\log(\mathrm{H_2O})$, and $\log(\mathrm{CO})$, with the standard and gravity-corrected retrievals shown in orange and blue, respectively. Solid vertical lines indicate the posterior medians, while horizontal error bars denote the corresponding $1\sigma$ credible intervals. The retrieved parameter values are listed above each posterior panel, with the standard and gravity-corrected results given from left to right.}}
    \label{fig:wasp-12bsynrectest}
\end{figure*}

{After generating the synthetic transmission spectrum including the centrifugal and tidal gravity corrections, atmospheric parameters are retrieved twice using identical atmospheric models: first with the standard formulation neglecting the gravity correction, and then with the gravity-corrected formulation. The adopted priors, injected atmospheric properties, and retrieved parameter estimates for both cases are listed in Table~\ref{tab:WASP12b_synthetic}. The corresponding retrieval results are shown in Figure~\ref{fig:wasp-12bsynrectest}.}

\begin{table}
\centering
\caption{{Atmospheric parameters adopted for the WASP-12b synthetic injection-recovery experiment. The synthetic spectrum was generated using the true atmospheric parameters listed below. Retrievals were subsequently performed using both the gravity-corrected and standard formulations. Quoted uncertainties correspond to the 1 $\sigma$ credible intervals.}}
\label{tab:WASP12b_synthetic}
\small
\setlength{\tabcolsep}{2.0pt}
\begin{tabular}{lccccc}
\hline
Parameter & Prior type & Prior range & True &
Retrieved value  &
Retrieved value  \\
 & & & value &
 (g-corrected) & (standard) \\
\hline

$\log(\mathrm{H_2O})$ &
Log-uniform &
$-12,\,-0.5$ &
$-2.00$ &
$-2.00^{+0.02}_{-0.02}$ &
$-2.14^{+0.02}_{-0.02}$ \\

$\log(\mathrm{CO})$ &
Log-uniform &
$-12,\,-0.5$ &
$-4.00$ &
$-4.23^{+0.50}_{-3.01}$ &
$-6.27^{+1.78}_{-3.94}$ \\

$T_{\rm iso}$ (K) &
Uniform &
$1600,\,3500$ &
2500 &
$2491^{+5.7}_{-5.9}$ &
$2832^{+2.7}_{-8.4}$ \\

$R_{\rm p,ref}$ ($R_{\rm J}$) &
Uniform &
$1.67,\,2.26$ &
1.965 &
$1.9649^{+0.0005}_{-0.0005}$ &
$1.9715^{+0.0006}_{-0.0004}$ \\

\hline
\end{tabular}
\end{table}







{The upper panel of Figure~\ref{fig:wasp-12bsynrectest} compares the retrieved transmission spectra with the simulated observations. The gravity-corrected retrieval accurately reproduces the injected spectrum, while the standard retrieval also provides an apparently satisfactory fit despite neglecting the reduction in effective gravity. The lower panels show the marginalized posterior distributions for the retrieved temperature and molecular abundances, allowing a direct comparison between the standard and gravity-corrected retrievals.}

{The gravity-corrected retrieval successfully recovers the injected atmospheric properties, yielding $T = 2491^{+6}_{-6}$~K and $\log(\mathrm{H_2O}) = -2.00^{+2}_{-2}$, in excellent agreement with the injected values of 2500~K and $\log(\mathrm{H_2O})=-2$. In contrast, the standard retrieval systematically compensates for the neglected gravity reduction by favouring a hotter atmosphere, recovering $T = 2832^{+3}_{-8}$~K, corresponding to an overestimation of approximately 340~K. Likewise, the retrieved water abundance shifts from the injected value of $\log(\mathrm{H_2O})=-2.00$ to $-2.14$, representing a decrease of approximately 0.14 dex.}

{The effect is considerably stronger for CO. While the gravity-corrected retrieval recovers $\log(\mathrm{CO})=-4.23$, close to the injected abundance of $-4$, the standard retrieval favours a substantially lower abundance of $\log(\mathrm{CO})=-6.27$, corresponding to a shift of approximately 2 dex. The posterior distribution of CO also becomes strongly asymmetric in the standard retrieval, exhibiting pronounced skewness toward lower abundances. This behaviour indicates that, in the absence of the gravity correction, the retrieval attempts to compensate for the underestimated atmospheric scale height by reducing the molecular abundances while simultaneously increasing the atmospheric temperature.}

{Although both retrievals reproduce the simulated transmission spectrum with comparable quality, only the gravity-corrected retrieval accurately recovers the injected atmospheric properties. The standard retrieval instead produces systematic biases in both the retrieved temperature and molecular abundances, demonstrating that neglecting centrifugal and tidal gravity corrections can lead to physically incorrect atmospheric inferences even when the spectral fit itself appears satisfactory.}

{\subsubsection{WASP-121b}}
For the WASP-121b analogue, we consider a cloud-free, isothermal atmosphere with an injected temperature of $T = 2400$~K and five spectroscopically active species: H$_2$O, CO, SiO, TiO and VO. The adopted injected abundances are $\log(X_{\mathrm{H_2O}}) = -5$, $\log(X_{\mathrm{CO}}) = -1$, $\log(X_{\mathrm{SiO}}) = -4$, $\log(X_{\mathrm{TiO}}) = -7$, and $\log(X_{\mathrm{VO}}) = -9$. {Synthetic spectra are generated over the wavelength range 0.6--5.2~$\mu$m, corresponding to the spectral coverage taken from \cite{Saha_2025}, and sampled into 214 spectral points with an assumed uncertainty of 30~ppm. This more complex atmospheric composition is chosen to resemble the chemical complexity inferred for ultra-hot Jupiters from recent JWST observations from \cite{Saha_2025} and provides a more realistic test of the gravity correction in the presence of multiple molecular opacity sources.}

\begin{figure*}
	\includegraphics[width=0.95\textwidth]{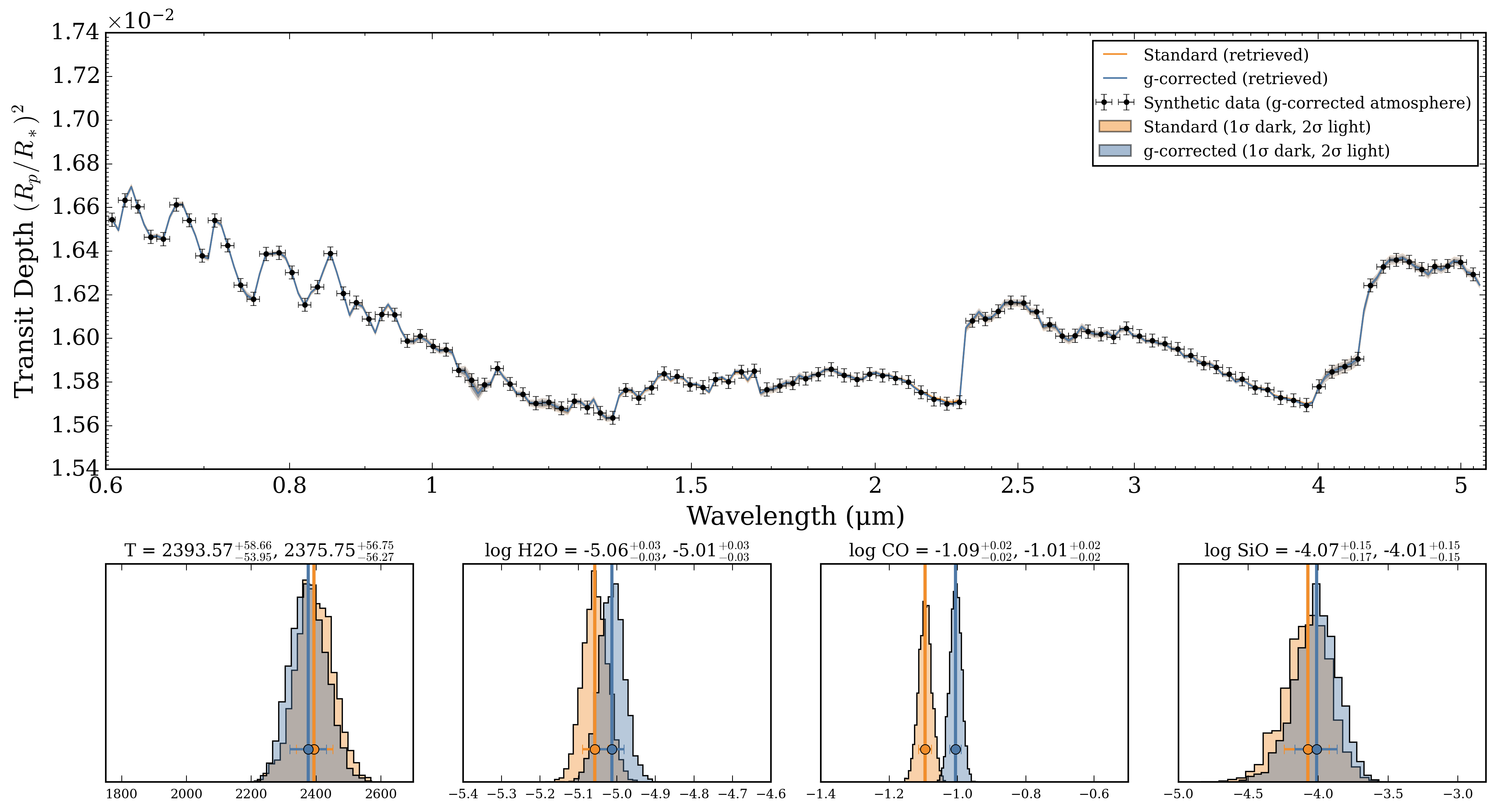}
    \caption{{Same as Figure~\ref{fig:wasp-12bsynrectest} but for WASP-121b. The upper panel compares the median retrieved transmission spectra obtained using the standard-gravity formulation ($g = 9.5565~\mathrm{m\,s^{-2}}$; orange) and the gravity-corrected formulation ($g' = 8.6275~\mathrm{m\,s^{-2}}$; blue). The lower panels show the corresponding marginal posterior distributions for the retrieved atmospheric temperature, $\log(\mathrm{H_2O})$, $\log(\mathrm{CO})$, and $\log(\mathrm{SiO})$, with the standard and gravity-corrected retrievals shown in orange and blue, respectively. Solid vertical lines indicate the posterior medians, while horizontal error bars denote the corresponding $1\sigma$ credible intervals. The retrieved parameter values are listed above each posterior panel, with the standard and gravity-corrected results given from left to right.}}
    \label{fig:wasp-121bsynrectest}
\end{figure*}

{Unlike the simplified WASP-12b case, the atmosphere with
five spectroscopically active species (H$_2$O, CO, SiO, TiO and VO),
allows us to investigate how the gravity correction propagates through
a higher-dimensional retrieval problem representative of recent
JWST-quality observations. The adopted atmospheric parameters and
retrieval results are summarized in Table~\ref{tab:wasp121b_synthetic},
while the corresponding posterior distributions are shown in
Figure~\ref{fig:wasp-121bsynrectest}.
}

\begin{table}
\centering
\caption{{Atmospheric parameters adopted for the WASP-121b synthetic injection-recovery experiment. Quoted uncertainties correspond to the $1\sigma$ credible intervals.}}
\label{tab:wasp121b_synthetic}
\small
\setlength{\tabcolsep}{2.0pt}
\begin{tabular}{lccccc}
\hline
Parameter & Prior type & Prior range & True &
Retrieved value &
Retrieved value \\
 & & & value &
 (g-corrected) & (standard) \\
\hline

$\log(\mathrm{H_2O})$
& Log-uniform
& $-12,\,-0.5$
& $-5.00$
& $-5.01^{+0.03}_{-0.03}$
& $-5.06^{+0.06}_{-0.06}$ \\

$\log(\mathrm{CO})$
& Log-uniform
& $-12,\,-0.5$
& $-1.00$
& $-1.01^{+0.02}_{-0.02}$
& $-1.10^{+0.04}_{-0.04}$ \\

$\log(\mathrm{SiO})$
& Log-uniform
& $-12,\,-0.5$
& $-4.00$
& $-4.01^{+0.15}_{-0.16}$
& $-4.07^{+0.30}_{-0.35}$ \\

$\log(\mathrm{TiO})$
& Log-uniform
& $-12,\,-0.5$
& $-7.00$
& $-7.00^{+0.03}_{-0.03}$
& $-7.05^{+0.06}_{-0.06}$ \\

$\log(\mathrm{VO})$
& Log-uniform
& $-12,\,-0.5$
& $-9.00$
& $-9.22^{+0.27}_{-0.62}$
& $-9.19^{+0.39}_{-2.33}$ \\

$T_{\rm iso}$ (K)
& Uniform
& $1800,\,3500$
& 2400
& $2375^{+57}_{-56}$
& $2392^{+59}_{-54}$ \\

$R_{\rm p,ref}$ ($R_{\rm J}$)
& Uniform
& $1.48,\,2.00$
& 1.742
& $1.743^{+0.002}_{-0.002}$
& $1.743^{+0.001}_{-0.001}$ \\

\hline
\end{tabular}
\end{table}

{The upper panel of Figure~\ref{fig:wasp-121bsynrectest} compares the
retrieved transmission spectra with the simulated observations.
Both the standard and gravity-corrected retrievals provide excellent
fits to the synthetic spectrum, with no visually distinguishable
difference between the retrieved spectra. The lower panels show the
marginalized posterior distributions for the retrieved temperature,
H$_2$O, CO, and SiO abundance. In contrast to the WASP-12b
experiment, the posterior distributions remain nearly Gaussian and
highly overlapping, with no evidence for the strong skewness or large
systematic offsets observed in the simpler two-species retrieval.}

{The gravity-corrected retrieval accurately recovers the injected
atmospheric state, yielding
$T=2375^{+57}_{-56}$ K,
$\log(\mathrm{H_2O})=-5.01^{+0.03}_{-0.03}$, $\log(\mathrm{CO})=-1.01^{+0.02}_{-0.02}$, $\log(\mathrm{SiO})=-4.01^{+0.15}_{-0.16}$, and 
$\log(\mathrm{TiO})=-7.00^{+0.02}_{-0.03}$,
in excellent agreement with the injected values of
2400 K,
$\log(\mathrm{H_2O})=-5$,
$\log(\mathrm{CO})=-1$, $\log(\mathrm{SiO})=-4$, and $\log(\mathrm{TiO})=-7$ .
The remaining molecular abundances are likewise recovered within their
$1\sigma$ uncertainties, with
$\log(\mathrm{VO})=-9.22^{+0.27}_{-0.62}$,
consistent with the injected abundances.}

When the same synthetic spectrum is analysed using the standard gravity retrieval, the recovered atmospheric parameters remain close to the injected values. {The retrieved temperature increases slightly to 
$2392^{+59}_{-54}$ K,
while the inferred abundances shift by only
$\sim0.05$ dex for H$_2$O,
$\sim0.09$ dex for CO,
$\sim0.06$ dex for SiO,
and remain statistically consistent with the injected values within the
retrieval uncertainties.}
Unlike the WASP-12b experiment, no pronounced skewness or strongly
biased posterior distributions are observed.

{Please note that we also performed a synthetic injection--recovery test for a simplified atmosphere containing only H$_2$O and CO$_2$, with injected abundances of $\log(X_{\mathrm{H_2O}})=-2.00$ and $\log(X_{\mathrm{CO_2}})=-4.00$. The gravity-corrected retrieval accurately recovered the injected atmosphere, yielding $\log(X_{\mathrm{H_2O}})=-2.00^{+0.04}_{-0.05}$ and $\log(X_{\mathrm{CO_2}})=-4.00^{+0.06}_{-0.07}$. In contrast, the standard-gravity retrieval recovered $\log(X_{\mathrm{H_2O}})=-2.26^{+0.06}_{-0.06}$ and $\log(X_{\mathrm{CO_2}})=-4.24^{+0.07}_{-0.06}$, corresponding to systematic offsets of approximately 0.26 dex and 0.24 dex, respectively, showing pronounced skewness. We do not discuss this test in detail because CO$_2$ has not been robustly detected in the available transmission spectrum of WASP-121b, and our primary synthetic analysis focuses on atmospheres containing molecules that have been observationally detected for this planet.}

{The comparison between the two synthetic experiments demonstrates that
the systematic biases introduced by neglecting the reduction in
effective gravity become substantially smaller when a chemically richer atmospheric model is
considered.
In this case, the additional molecular opacity sources provide greater
freedom for the retrieval to reproduce the transmission spectrum,
resulting in only modest differences between the standard and
gravity-corrected solutions.
Nevertheless, the gravity-corrected retrieval consistently recovers the
injected atmospheric parameters more accurately, confirming that the
effective reduction in gravity remains detectable even in a more
realistic JWST retrieval scenario.}

{The synthetic injection-recovery experiments presented above provide a controlled assessment of how centrifugal and tidal gravity corrections propagate into retrieved atmospheric parameters. Having verified the behaviour of the retrieval framework for both a simplified atmosphere (WASP-12b) and a more chemically complex JWST analogue (WASP-121b), we now apply the same retrieval methodology to observed transmission spectra. Two independent retrievals are performed for each planet: one using the standard formulation and the other including the effective gravity correction described in Section~2.3. We first present the retrieval results for the HST/WFC3 G141 transmission spectrum of WASP-12b, followed by the higher-precision JWST/NIRSpec G395H observations of WASP-121b, enabling a direct comparison of the impact of centrifugal and tidal gravity corrections on atmospheric inferences from current observational datasets.}

\subsection{Retrieved atmospheric abundances from real observations}
We perform atmospheric retrievals on the observed transmission spectra of WASP-12b obtained with \textit{HST}/WFC3 G141 \citep{2022ApJS..260....3C} and WASP-121b observed with JWST/NIRSpec G395H \citep{Gapp_2025}. Uniform priors are adopted for all retrieved parameters, including the atmospheric temperature, reference radius, and volume mixing ratios of the molecular species included in each model.

\subsubsection{WASP-12b}
\label{sec:retrieval}

\begin{figure*}
	\includegraphics[width=0.95\textwidth]{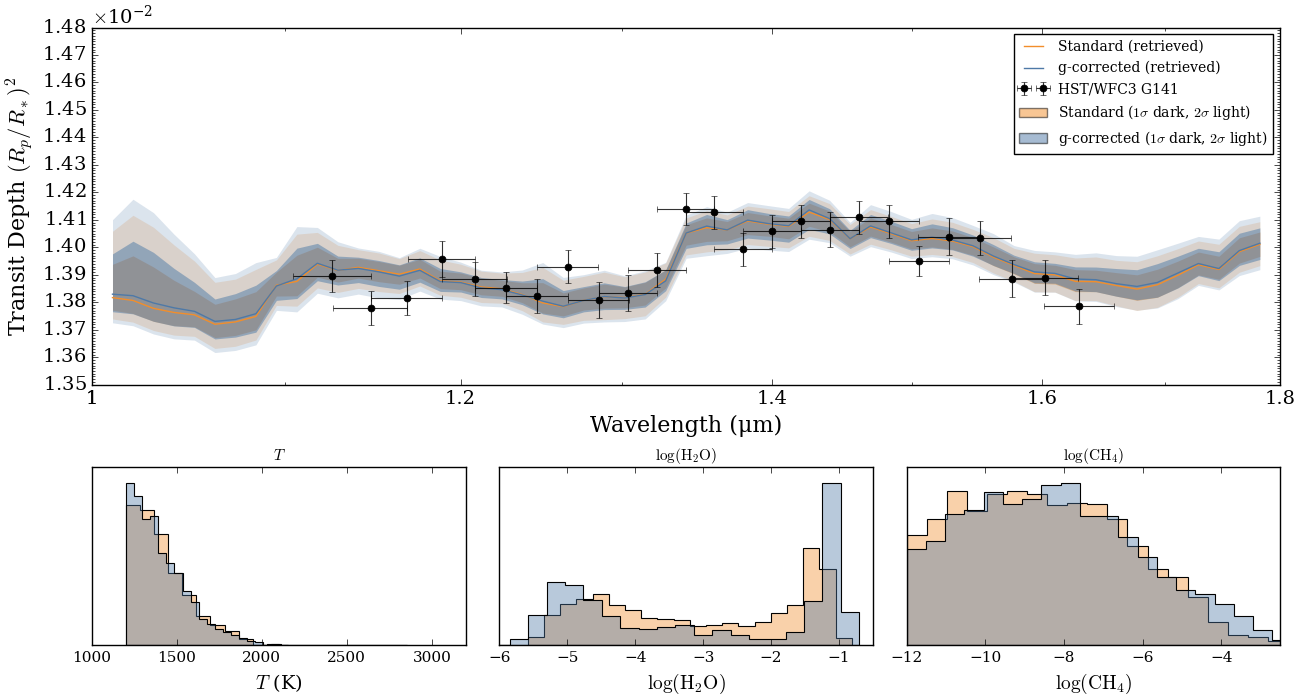}
    \caption{
Retrieved transmission spectrum and posterior distributions for the isothermal, cloud-free model of WASP-12b using HST/WFC3 data, including a gravity-correction scenario. The top panel shows the median retrieved spectrum, with shaded regions indicating the 1$\sigma$ and 2$\sigma$ credible intervals. Orange and blue curves correspond to the standard ($g = 9.4362\,\mathrm{m\,s^{-2}}$) and gravity-corrected cases ($g' = 7.9447\,\mathrm{m\,s^{-2}}$), respectively. The lower panels display marginalized posterior distributions for selected parameters.The $H_2O$ posterior is bimodal; therefore median and 1$\sigma$ summary statistics are not shown for this parameter.
}
    \label{fig:wasp-12bretrieval}
\end{figure*}

{We use the available HST/WFC3 G141 transmission spectrum of WASP-12b \citep{2022ApJS..260....3C} to perform retrievals for both the standard and gravity-corrected formulations. The retrieval configuration follows an isothermal, cloud-free atmosphere with H$_2$O, CH$_4$, CO, CO$_2$, NH$_3$, and TiO as free molecular species. The adopted prior distributions together with the retrieved atmospheric parameters are summarized in Table~\ref{tab:wasp12b_retrieval}, while the corresponding retrieved spectra and posterior distributions are shown in Figure~\ref{fig:wasp-12bretrieval}.
}
{Figure~\ref{fig:wasp-12bretrieval} compares the retrieved transmission spectra with the observed HST/WFC3 G141 data. Both the standard and gravity-corrected retrievals reproduce the observed transmission spectrum equally well, yielding nearly indistinguishable median spectra within the observational uncertainties. The lower panels display the marginalized posterior distributions for representative atmospheric parameters, allowing a direct comparison between the two retrievals.}

{Unlike the controlled synthetic injection--recovery experiments presented in Section~3.2, the atmospheric retrievals using the HST/WFC3 G141 transmission spectrum of WASP-12b provide only weak observational constraints on the effect of centrifugal acceleration and tidal gravity. The most notable feature is the posterior distribution of H$_2$O, which exhibits a distinctly bimodal structure in both the standard and gravity-corrected retrievals. Since the posterior is strongly non-Gaussian, its median and corresponding $1\sigma$ credible interval do not provide a complete summary of the distribution. We therefore compare the full posterior distributions in Figure~\ref{fig:wasp-12bretrieval} rather than interpreting the H$_2$O medians alone.}

{The remaining retrieved parameters show broad posterior distributions with substantial overlap between the standard and gravity-corrected retrievals. The largest shifts occur for TiO, whose posterior medians move by approximately 0.3 dex, while CH$_4$ and NH$_3$ exhibit comparatively smaller offsets. In contrast, the retrieved planetary radius, atmospheric temperature, CO and CO$_2$ abundance remain consistent within their quoted $1\sigma$ credible intervals.}

These broad posterior distributions primarily reflect the limited wavelength coverage, spectral resolution, and signal-to-noise ratio of the HST/WFC3 G141 observations. Consequently, although the retrievals show systematic shifts after incorporating the reduced effective gravity, the HST data do not provide statistically significant evidence for gravity-induced changes in the atmospheric abundances. {Instead, the WASP-12b retrieval from observations serves as a supporting case, while the controlled synthetic injection-recovery experiments provide a framework for assessing how neglecting the gravity correction can bias the retrieved atmospheric parameters.}

\begin{table}
\centering
\small
\setlength{\tabcolsep}{2.5pt}

\caption{Atmospheric retrieval results for WASP-12b using HST/WFC3 G141 data, comparing the standard-gravity and gravity-corrected retrievals including centrifugal and tidal effects. Uncertainties correspond to the $1\sigma$ credible intervals.}
\label{tab:wasp12b_retrieval}
\small
\setlength{\tabcolsep}{1.5pt}
\begin{tabular}{lcccc}
\hline
Parameter & Prior type & Prior range &
Retrieved value & Retrieved value \\
& & & 
 (standard) & (g-corrected) \\
\hline

$R_{\rm p,ref}$ ($R_{\rm J}$) &
Uniform &
$1.67,\,2.26$ &
$1.89^{+0.02}_{-0.01}$ &
$1.88^{+0.02}_{-0.02}$ \\

$T$ (K) &
Uniform &
$1000,\,3200$ &
$1375.3^{+205.6}_{-119.3}$ &
$1336.3^{+203.1}_{-120.5}$ \\

$\log(\mathrm{H_2O})$ &
Log-uniform &
$-12,\,-0.5$ &
$-2.98^{+1.64}_{-1.72}$ &
$-3.16^{+2.08}_{-1.89}$ \\

$\log(\mathrm{CH_4})$ &
Log-uniform &
$-12,\,-0.5$ &
$-8.58^{+2.35}_{-2.20}$ &
$-8.31^{+2.45}_{-2.28}$ \\

$\log(\mathrm{CO})$ &
Log-uniform &
$-12,\,-0.5$ &
$-7.58^{+3.35}_{-2.85}$ &
$-7.50^{+2.87}_{-2.90}$ \\

$\log(\mathrm{CO_2})$ &
Log-uniform &
$-12,\,-0.5$ &
$-7.78^{+2.87}_{-2.72}$ &
$-7.84^{+2.82}_{-2.55}$ \\

$\log(\mathrm{NH_3})$ &
Log-uniform &
$-12,\,-0.5$ &
$-6.88^{+2.43}_{-3.36}$ &
$-6.70^{+2.31}_{-3.24}$ \\

$\log(\mathrm{TiO})$ &
Log-uniform &
$-12,\,-0.5$ &
$-8.10^{+2.13}_{-2.37}$ &
$-7.76^{+2.37}_{-2.60}$ \\

\hline
\end{tabular}
\end{table}

{\subsubsection{WASP-121b}}

\begin{figure*}
	\includegraphics[width=0.95\textwidth]{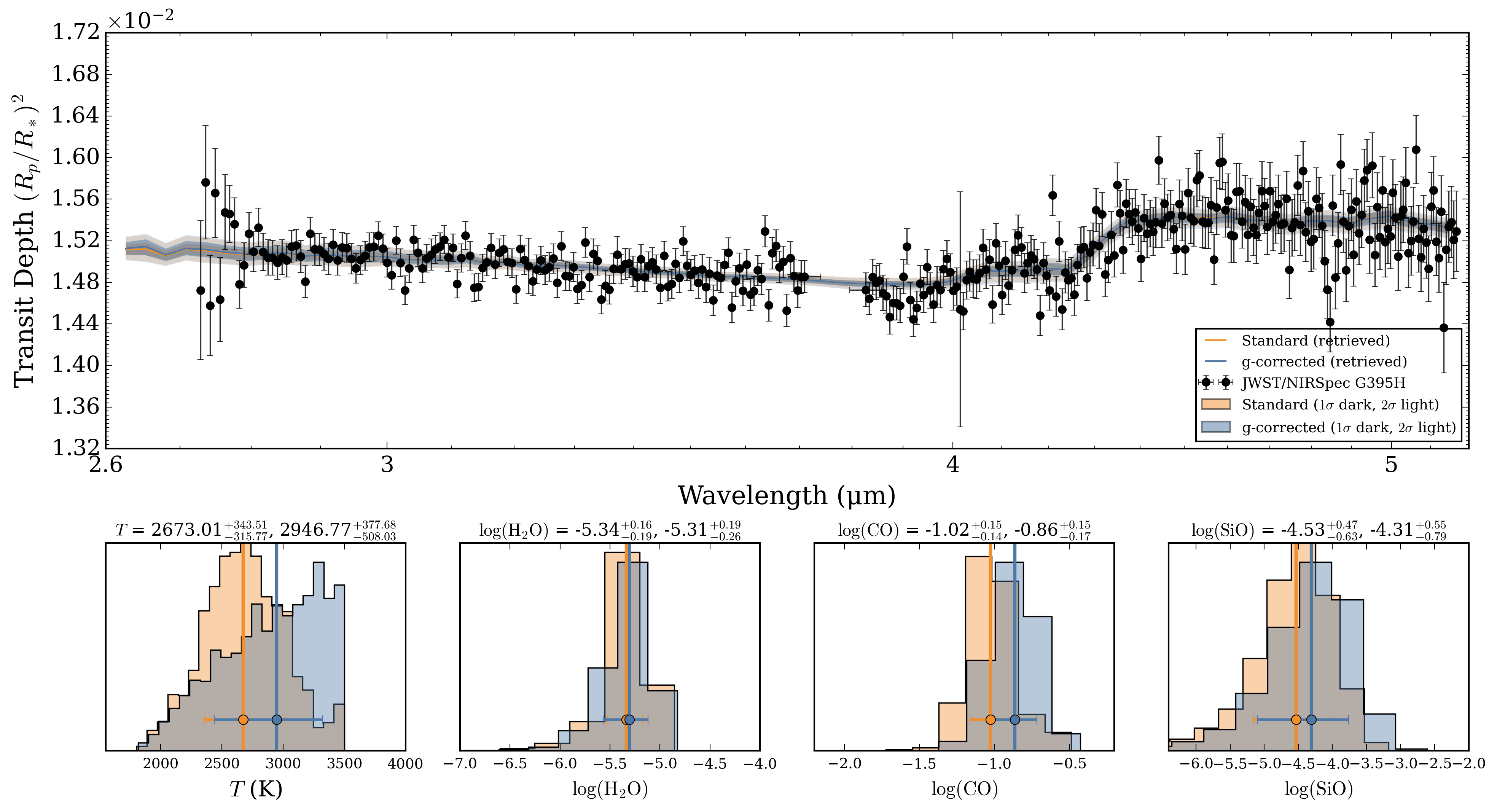}
    \caption{
{Same as Figure~\ref{fig:wasp-12bretrieval} but for WASP-121b using JWST/NIRSpec G395H data. The top panel shows the median retrieved spectrum, with shaded regions indicating the 1$\sigma$ and 2$\sigma$ credible intervals. Orange and blue curves correspond to the standard ($g = 9.5565\,\mathrm{m\,s^{-2}}$) and gravity-corrected cases ($g' = 8.6275\,\mathrm{m\,s^{-2}}$), respectively. The lower panels display marginalized posterior distributions for selected parameters. Solid vertical lines denote the median values, while horizontal error bars indicate the corresponding 1$\sigma$ credible intervals. The retrieved parameter values are listed above each panel, with the standard (orange) and gravity-corrected (blue) cases shown from left to right.
}}
    \label{fig:wasp-121bretrieval}
\end{figure*}

We next perform atmospheric retrievals for WASP-121b using the
JWST/NIRSpec G395H transmission spectrum. Two retrievals are carried
out using identical atmospheric models and prior distributions, differing
only in the treatment of the effective gravitational acceleration. {
Both retrievals assume an isothermal, cloud-free atmosphere containing
H$_2$O, CO, and SiO as free molecular species taken from \cite{evanssoma2025siosuperstellarcoratio}. We omit the inclusion of TiO and VO as free parameters as they lie outside of the wavelength range of JWST/NIRSpec G395H data and are not sufficiently constrained from the same.} The adopted priors and retrieved atmospheric parameters are summarized in
Table~\ref{tab:wasp121b_retrieval}, while the corresponding retrieved spectra and posterior distributions are shown in
Figure~\ref{fig:wasp-121bretrieval}.

\begin{table}
\centering
\caption{{Atmospheric retrieval results for WASP-121b using JWST/NIRSpec G395H transmission spectroscopy, comparing the standard gravity formulation with the gravity-corrected model including centrifugal and tidal effects. Uncertainties correspond to the $1\sigma$ credible intervals.}}
\label{tab:wasp121b_retrieval}
\small
\setlength{\tabcolsep}{1.5pt}
\begin{tabular}{lcccc}
\hline
Parameter & Prior type & Prior range &
Retrieved value &
Retrieved value \\
& & & 
 (standard) & (g-corrected) \\
\hline

$R_{\rm p,ref}\,(R_{\rm J})$
& Uniform
& $1.48,\,2.00$
& $1.69^{+0.01}_{-0.01}$
& $1.69^{+0.01}_{-0.01}$ \\

$T$ (K)
& Uniform
& $1800,\,3500$
& $2673.2^{+349.0}_{-317.5}$
& $2950.2^{+373.6}_{-511.1}$ \\

$\log(\mathrm{H_2O})$
& Log-uniform
& $-12,\,-0.1$
& $-5.34^{+0.16}_{-0.20}$
& $-5.31^{+0.19}_{-0.26}$ \\

$\log(\mathrm{CO})$
& Log-uniform
& $-12,\,-0.1$
& $-1.02^{+0.15}_{-0.14}$
& $-0.86^{+0.15}_{-0.17}$ \\

$\log(\mathrm{SiO})$
& Log-uniform
& $-12,\,-0.1$
& $-4.53^{+0.48}_{-0.65}$
& $-4.32^{+0.55}_{-0.78}$ \\



\hline
\end{tabular}
\end{table}

{The upper panel of Figure~\ref{fig:wasp-121bretrieval} compares the
retrieved transmission spectra with the observed JWST/NIRSpec G395H
data. Both the standard and gravity-corrected retrievals reproduce the
observed spectrum with comparable quality, yielding nearly identical
median spectra and reduced chi-square values of
$\chi_{\rm red}^2 \simeq 1.00$. 
The lower panels display the marginalized posterior distributions for
the atmospheric temperature together with representative molecular
abundances, illustrating the effect of including the centrifugal and
tidal gravity corrections during the retrieval.}

The retrieved atmospheric parameters remain broadly consistent between
the two retrievals. {The gravity-corrected model yields
$T = 2950^{+374}_{-511}$ K, while the corresponding standard retrieval
returns $T = 2673^{+349}_{-317}$ K, corresponding to a difference of
approximately 277 K. Likewise, the retrieved molecular abundances show
only modest offsets, with
$\Delta\log(\mathrm{H_2O}) \approx 0.03$ dex,
$\Delta\log(\mathrm{CO}) \approx 0.16$ dex, and 
$\Delta\log(\mathrm{SiO}) \approx 0.21$ dex
between the standard and gravity-corrected solutions.} The posterior
distributions remain approximately Gaussian for both retrievals, with
substantial overlap between the two cases and no evidence for the
multimodal behaviour encountered in the earlier HST/WFC3 retrievals.

{Compared with the synthetic injection-recovery experiments, the retrievals using the observed JWST spectrum exhibit relatively small
parameter shifts while maintaining an excellent fit to the data. The
results therefore indicate that incorporating centrifugal and tidal
gravity corrections produces systematic changes in the retrieved
atmospheric properties, although the magnitude of these changes remains
modest for the present JWST observations.}


{The list of planet candidates presented in Table~\ref{tab:target} indicates that the combined effects of centrifugal acceleration and stellar tides can reduce the effective surface gravity of the most strongly affected close-in giant planets by up to $\sim16\%$. Although systems such as WASP-12b represent this extreme regime, high-precision JWST transmission spectra are not currently available for these targets. We therefore perform an additional retrieval experiment for WASP-121b by adopting a $16\%$ reduction in the effective gravity, allowing us to investigate how retrieval results change under the maximum gravity corrections using JWST-quality observations.}

We adopt the same cloud-free, isothermal atmospheric model and identical priors, modifying only the effective gravity by $16\%$. The retrieved atmospheric parameters are listed in Table~\ref{tab:wasp121_16}. Overall, the inferred atmospheric properties exhibit only modest changes compared to the nominal $10\%$ gravity-corrected retrieval discussed previously. {The retrieved temperature increases from $2673$~K to $2966$~K, while the H$_2$O abundance changes from $\log(X_{\mathrm{H_2O}})=-5.34$ to $-5.30$. Among the molecular species, CO shows the largest shift, increasing from $\log(X_{\mathrm{CO}})=-1.02$ to $-0.80$, whereas SiO increases from $\log(X_{\mathrm{SiO}})=-4.53$ to $-4.31$. These results suggest that increasing the gravity correction from $10\%$ to the maximum value of approximately $16\%$ does not alter the retrieved atmospheric parameters significantly, indicating similar (as in real 10\% gravity reduction case) systematic trend in retrieved parameters.}

\begin{table}
\centering
\caption{{Retrieved atmospheric parameters for the additional WASP-121b retrieval assuming a $16\%$ reduction in effective gravity.}}
\label{tab:wasp121_16}
\begin{tabular}{lcc}
\hline
Parameter & Retrieved value & Retrieved value \\
& (standard) & (g-corrected, 16\%)\\
\hline
$T$ (K) & $2673.2^{+349.0}_{-317.5}$ & $2965.9^{+384.6}_{-546.2}$\\
$\log(\mathrm{H_2O})$ & $-5.34^{+0.16}_{-0.20}$ & $-5.30^{+0.21}_{-0.31}$\\
$\log(\mathrm{CO})$ & $-1.02^{+0.15}_{-0.14}$ & $-0.80^{+0.16}_{-0.17}$\\
$\log(\mathrm{SiO})$ & $-4.53^{+0.48}_{-0.65}$ & $-4.31^{+0.51}_{-0.72}$\\
\hline
\end{tabular}
\end{table}

\begin{table*}
\centering
\caption{Retrieved atmospheric parameters for {WASP-121b} under different model assumptions. Quoted uncertainties correspond to the $1\sigma$ credible intervals.}
\label{tab:wasp121b_retrieval_diff_model}
\resizebox{\textwidth}{!}{
\begin{tabular}{lcccccc}
\hline\hline

& \multicolumn{2}{c}{Simple model (Isothermal)}
& \multicolumn{2}{c}{Non-isothermal P-T profile \citep{Madhusudhan_2009}}
& \multicolumn{2}{c}{Cloudy model \citep{MacDonald_2017}} \\

\cline{2-7}

Parameter
& Standard
& gravity-corrected
& Standard
& gravity-corrected
& Standard
& gravity-corrected \\

\hline

$R_{\rm p,ref}$ ($R_{\rm J}$)
&
$1.69^{+0.01}_{-0.01}$
&
$1.69^{+0.01}_{-0.01}$
&
$1.69^{+0.01}_{-0.01}$
&
$1.69^{+0.01}_{-0.01}$
&
$1.69^{+0.01}_{-0.01}$
&
$1.69^{+0.01}_{-0.01}$ \\

$T$ (K)
&
$2673^{+349}_{-317}$
&
$2950^{+374}_{-511}$
&
---
&
---
&
$2943^{+367}_{-487}$
&
$2878^{+382}_{-457}$ \\

$T_{\rm ref}$ (K)
&
---
&
---
&
$2897^{+340}_{-498}$
&
$2887^{+376}_{-447}$
&
---
&
--- \\

\hline

$a_1$
&
---
&
---
&
$0.52^{+0.26}_{-0.24}$
&
$0.61^{+0.25}_{-0.28}$
&
---
&
--- \\

$a_2$
&
---
&
---
&
$0.60^{+0.23}_{-0.27}$
&
$0.58^{+0.25}_{-0.26}$
&
---
&
--- \\

$\log P_1$
&
---
&
---
&
$-2.43^{+1.72}_{-1.52}$
&
$-2.78^{+1.86}_{-1.51}$
&
---
&
--- \\

$\log P_2$
&
---
&
---
&
$-2.42^{+1.81}_{-1.61}$
&
$-2.33^{+1.80}_{-1.58}$
&
---
&
--- \\

$\log P_3$
&
---
&
---
&
$0.73^{+0.77}_{-1.19}$
&
$0.48^{+0.99}_{-1.33}$
&
---
&
--- \\

\hline

$\log X_{\rm H_2O}$
&
$-5.34^{+0.16}_{-0.20}$
&
$-5.31^{+0.19}_{-0.26}$
&
$-5.33^{+0.15}_{-0.20}$
&
$-5.33^{+0.17}_{-0.24}$
&
$-5.33^{+0.18}_{-0.24}$
&
$-5.32^{+0.18}_{-0.25}$ \\

$\log X_{\rm CO}$
&
$-1.02^{+0.15}_{-0.14}$
&
$-0.86^{+0.15}_{-0.17}$
&
$-0.97^{+0.13}_{-0.19}$
&
$-0.87^{+0.15}_{-0.15}$
&
$-0.95^{+0.16}_{-0.19}$
&
$-0.87^{+0.16}_{-0.17}$ \\

$\log X_{\rm SiO}$
&
$-4.53^{+0.48}_{-0.65}$
&
$-4.32^{+0.55}_{-0.78}$
&
$-4.35^{+0.41}_{-0.63}$
&
$-4.35^{+0.48}_{-0.78}$
&
$-4.36^{+0.48}_{-0.78}$
&
$-4.34^{+0.47}_{-0.81}$ \\



$\log P_{\rm cloud}$
&
---
&
---
&
---
&
---
&
$0.44^{+1.04}_{-1.08}$
&
$0.51^{+0.95}_{-1.11}$ \\

\hline
\end{tabular}
}
\end{table*}

\subsection{Robustness of retrieved atmospheric abundances under gravity correction}
All the retrieval results presented so far have assumed an isothermal and could free atmosphere. In this section, we evaluate whether non-isothermal atmosphere and presence of clouds can affect the inferred atmospheric parameters by capping the effect of rotation and stellar tide on the planetary atmosphere. 

{\subsubsection{Effect of gravity-correction on a non-isothermal atmosphere}}\label{subsubsection:madhu}
{The non-isothermal temperature structure is described using the parametric
$P$-$T$ profile of \citet{Madhusudhan_2009}, which divides the atmosphere
into three pressure regimes. The upper two layers are described by generalized
exponential profiles,}
{\begin{equation}
P=P_0\exp\left[a_1(T-T_0)^{1/2}\right],
\end{equation}
and
\begin{equation}
P=P_2\exp\left[a_2(T-T_2)^{1/2}\right],
\end{equation}
}
{while the deepest layer is assumed to be isothermal, $T=T_3$. Here, $a_1$ and $a_2$ control the temperature gradients in Layers 1 and 2, respectively. $T_0$ and $T_1$ are the free parameters. The parameters $\log P_1$, $\log P_2$, and $\log P_3$ specify characteristic
pressure levels in the profile, with $P_1$ and $P_3$ marking the transitions
between the three layers, while $P_2$ is the reference pressure associated
with the Layer-2 profile. The pressures are expressed as
$\log_{10}(P/\mathrm{bar})$ \citep{Madhusudhan_2009}. In this parametrization, $P_2>P_1$ corresponds to a thermal inversion,
whereas $P_1\geq P_2$ describes a profile without a thermal inversion.}

When adopting the non-isothermal P-T profile of \citet{Madhusudhan_2009}, the number of free parameters increases substantially compared to the simple isothermal model, allowing for a more flexible representation of the atmospheric thermal structure. Despite this additional model complexity, the retrieved atmospheric parameters remain broadly consistent between the standard and gravity-corrected retrievals. {The inferred H$_2$O, CO, and SiO abundances show only modest shifts.} These results indicate that the inclusion of a flexible P-T profile does not qualitatively alter the impact of reduced effective gravity. Instead, the additional flexibility allows modest redistribution between thermal structure and composition.

{\subsubsection{Effect of gravity-correction on a cloudy atmosphere}}
\label{subsubsection:macmad17}
{To examine whether the inclusion of clouds affects the response of the retrieved atmospheric parameters to centrifugal and tidal gravity corrections, we additionally consider the grey cloud prescription of \citet{MacDonald_2017}. In transmission spectroscopy, an optically thick cloud deck can limit the pressure range accessible to observations by truncating the transmission spectrum at pressures below the cloud-top pressure. This reduces the effective atmospheric extent probed by the transmission spectrum and can therefore modify the sensitivity of molecular features to changes in effective gravity.}

{For the cloud model, we introduce the cloud-top pressure, $P_{\rm cloud}$, as an additional free parameter. The cloud deck is assumed to be opaque at pressures greater than $P_{\rm cloud}$, while the atmosphere above the cloud deck remains accessible to transmission spectroscopy. This provides a simple parameterization of the effect of clouds without requiring assumptions about the detailed chemical composition or particle properties of the cloud.}

{The effect of this additional atmospheric component is investigated using the JWST/NIRSpec G395H transmission spectrum of WASP-121b. We perform two otherwise identical retrievals, one using the standard effective gravity and the other incorporating the combined centrifugal and tidal gravity correction. The atmospheric parameters, priors, and remaining retrieval settings are kept consistent between the two cases, with $P_{\rm cloud}$ included as an additional free parameter. The retrieved parameters for the standard and gravity-corrected cases are presented together with the corresponding isothermal and non-isothermal retrievals in Table~\ref{tab:wasp121b_retrieval_diff_model}. For this case also, we see only modest differences between the standard and gravity-corrected retrievals. The retrieved molecular abundances remain consistent within the quoted uncertainties, while the cloud-top pressure itself changes only marginally between the two cases. Although clouds are expected to reduce the observable atmospheric scale height by truncating the transmission spectrum at higher pressures, our retrievals indicate that, for the present JWST observations of WASP-121b, the inclusion of a simple grey cloud deck does not significantly modify the inferred impact of reduced effective gravity. Overall, these experiments demonstrate that the qualitative influence of centrifugal acceleration and tidal forces remains robust against commonly adopted atmospheric model assumptions, including more flexible temperature structures and the presence of grey clouds.}\\

\section{Discussion and Conclusion}
In this work, we investigated the impact of centrifugal and tidal gravity corrections on atmospheric molecular abundances inferred from transmission spectra using retrieval techniques for close-in giant planets. To incorporate the combined effect of both rotational and tidal effects, we have formulated a consolidated way in which the effective gravitational acceleration for the planet has been reduced due to centrifugal and tidal forces. The tidal and centrifugal contributions in reducing the effective gravity have been calculated from the observed values of planetary rotation, orbital distance, and the mass of the host star. Once we estimate the reduced effective gravity in terms of the modified gravitational acceleration ($g'$), we perform the atmospheric retrieval using \texttt{POSEIDON} with this updated value. {Through its influence on the atmospheric scale height, the modified gravity changes the predicted transmission spectrum and can consequently alter the retrieved log-mixing ratios of molecular species in close-in exoplanets where centrifugal acceleration and tidal effects are non-negligible.} Before even moving towards the retrieval framework, we estimated the effect of reduced $g'$ in the transmission spectra using a forward model and found that tidal and rotation could modify the transmission spectra significantly. We presented two case studies using a forward model, where we showed that for a reduced effective gravity for WASP-12b, there are changes of {300-600} ppm in transit depth for different molecular species, whereas for the planet {WASP-121b} where the reduction in effective gravity is smaller, the transit spectra showed changes in transit depth from {60-140} ppm. As these changes are quite significant, we believe that the signatures induced by centrifugal acceleration and tidal forces in the atmosphere should, in principle, be detectable in transmission spectra. {To validate this hypothesis, we performed a series of synthetic injection-recovery experiments using \texttt{POSEIDON}, allowing us to investigate the impact of reduced effective gravity under controlled conditions where the true atmospheric state is assumed to be known.}

{The synthetic injection-recovery experiments proved particularly useful in distinguishing the physical impact of reduced effective gravity from retrieval degeneracies. For relatively simple atmospheric models containing only a small number of free parameters (e.g., H$_2$O and CO), neglecting the reduction in effective gravity leads to noticeably larger shifts in the retrieved molecular abundances, whereas retrievals accounting for the reduced effective gravity recover the injected atmospheric state more closely. As the atmospheric model becomes increasingly complex, with additional molecular species and therefore a larger parameter space, these shifts become comparatively smaller. In such cases, the retrieval is able to redistribute the spectral information among multiple atmospheric parameters, thereby reducing the apparent change in any individual molecular abundance. This behaviour is illustrated by the WASP-121b retrievals, where the inclusion of additional absorbers results in more modest differences between the standard and gravity-corrected retrievals despite the underlying physical modification to the atmospheric scale height. These synthetic experiments therefore provide two important insights. First, they demonstrate that the abundance shifts induced by reduced effective gravity are physical and can be recovered by atmospheric retrievals when the atmospheric state is well constrained. Second, they show that the magnitude of the retrieved shifts depends not only on the reduction in effective gravity itself, but also on the complexity of the adopted atmospheric model and the associated retrieval degeneracies. Consequently, the synthetic injection-recovery tests provide a good controlled framework for interpreting the results where differences obtained from retrievals of standard gravity and reduced gravity with real observational data turns out to be modest.}

Next, we performed atmospheric retrievals for WASP-12b (HST/WFC3 G141) and {WASP-121b (JWST/NIRSpec G395H)} with and without incorporating the modified effective gravity. {For WASP-12b, the limited wavelength coverage and comparatively lower precision of the HST observations result in retrievals that remain highly degenerate, making it difficult to isolate the effects of reduced gravity from other atmospheric parameters. In contrast, the higher spectral precision and broader wavelength coverage of the JWST observations for WASP-121b enable a more detailed investigation of the influence of effective gravity on the retrieved atmospheric properties. Although the changes in the retrieved molecular abundances and temperature are generally modest, the gravity-corrected retrievals consistently favour slightly different atmospheric solutions compared to the standard retrievals.}

{Motivated by the maximum observed reduction of gravity due to rotation and tidal force in the candidate planets shown in the Table~\ref{tab:target}), which for some close-in giant planets can be up to approximately $16\%$, we further performed an additional retrieval experiment for WASP-121b adopting this maximum gravity correction. This controlled experiment provides an opportunity to investigate the influence of stronger gravity reductions using JWST-quality observations, despite the absence of comparable JWST transmission spectra for the most strongly affected systems. The retrieved atmospheric parameters exhibit only slight additional changes relative to the nominal gravity-corrected retrieval, suggesting that increasing the effective gravity reduction from the typical value adopted in this work (~$10\%$) to the maximum physically expected value (~$16\%$) does not qualitatively alter the inferred atmospheric composition.} 

{Although the forward-model calculations and synthetic injection-recovery experiments demonstrate that reduced effective gravity can produce measurable changes in transmission spectra and the retrieved atmospheric properties, we emphasize that the corresponding retrievals using current observational data generally exhibit more modest differences. Nevertheless, while interpreting these results, we should note that we have not accounted for the latitudinal variation of the centrifugal and tidal contributions when performing our forward-model calculations and atmospheric retrievals. It has already been shown by \citet{Banerjee_2023} that incorporating the latitudinal dependence of the centrifugal acceleration introduces only minor variations in the transmission spectrum. They concluded that latitudinal variation in rotation would introduce a change of around 5 ppm in the transmission spectra. For the tidal part, we have not performed any separate calculation considering the latitudinal variation of tidal gravity. However, a future study should include the variation, we believe that the latitudinal variation would not contribute to the changes more significantly than the current estimation of equatorial tidal force from the forward model, as well as retrievals.}

In all of our retrieval studies, we have retrieved atmospheric parameters for close-in hot Jupiters using a simple, isothermal, cloud-free model. An important question is whether the inclusion of additional atmospheric complexities, such as a more flexible pressure-temperature (P-T) profile or the presence of clouds, modifies the impact of centrifugal and tidal
accelerations on the retrieved atmospheric parameters. To address this, we repeated the retrieval analysis for WASP-121b using two additional atmospheric models beyond the simple case: (i) a non-isothermal P-T profile following the parameterization of \citep{Madhusudhan_2009}, and (ii) a grey cloud deck using the prescription of \citep{MacDonald_2017}. The description of both these models are introduced in Section~3.4.1 and 3.4.2. All retrievals were performed using \textsc{POSEIDON} for both the standard-gravity and gravity-corrected cases. The retrieved parameter values for all three models, with and without gravity corrections, are summarized in Table~\ref{tab:wasp121b_retrieval_diff_model}. {Across the standard isothermal, Madhusudhan--Seager P--T profile, and grey cloud models, the retrieved H$_2$O abundance remains essentially unchanged, with values of $\log X_{\rm H_2O}=-5.34$ and $-5.31$, $-5.33$ and $-5.33$, and $-5.33$ and $-5.32$, respectively, for the standard and gravity-corrected retrievals. In contrast, CO consistently exhibits a systematic shift between the two retrievals across all three atmospheric models, changing from $\log X_{\rm CO}=-1.02$ to $-0.86$ (isothermal), $-0.97$ to $-0.87$ (Madhusudhan--Seager), and $-0.95$ to $-0.87$ (cloudy), demonstrating that the qualitative impact of the gravity correction is robust to the adopted atmospheric model assumptions.}

{In conclusion, close-in tidally locked giant planets can experience appreciable reductions in their effective surface gravity owing to the combined effects of centrifugal acceleration and stellar tides. These reductions increase the atmospheric scale height and produce measurable changes in transmission spectra, with forward-model calculations predicting transit-depth differences of up to several hundred ppm for strongly distorted systems. {Our synthetic injection-recovery experiments further demonstrate that neglecting the gravity correction can produce systematic biases in retrieved atmospheric parameters under controlled conditions, while also illustrating how the magnitude of the inferred abundance shifts depends on the complexity of the retrieval model and the number of free atmospheric parameters. Hence, although retrievals show a promising way to quantify the gravity correction, inferring the magnitude of the gravity correction on abundances would be difficult from the retrievals themselves, as the standard and gravity-corrected models can provide comparably good fits to observational data.} Applying our retrieval framework to HST observations of WASP-12b and JWST observations of WASP-121b, we find that the inclusion of effective gravity corrections generally leads to systematic, but comparatively modest, shifts in the retrieved atmospheric abundances. Although these differences are often comparable to the current observational uncertainties, they remain qualitatively consistent across different atmospheric assumptions, including non-isothermal pressure-temperature profiles and cloudy atmospheres. This suggests that the inferred influence of reduced effective gravity is robust and not an artifact of a particular retrieval parameterization. Finally, we have identified a population of highly distorted Gas-Giants (see Table~\ref{tab:target}) that experience substantial reductions in effective gravity. These systems represent the most promising targets for future high-precision spectroscopic observations aimed at directly probing the influence of centrifugal acceleration and stellar tides on exoplanet atmospheres. Incorporating these effects into atmospheric retrieval frameworks will therefore be essential for achieving physically self-consistent interpretations of transmission spectra with current and next-generation facilities.}

\section*{Acknowledgements}
We are grateful to the reviewer Joanna Barstow for a constructive report that significantly helped to improve the manuscript. The authors would also like to thank Nikku Madhusudhan, Liton Majumdar, Jayesh Goyal, Ryan MacDonald, and Vigneshwaran Krishnamurthy for many fruitful discussions that helped the presentation of our work. This research is based on observations made with the NASA/ESA Hubble Space Telescope obtained from the Space Telescope Science Institute, which is operated by the Association of Universities for Research in Astronomy, Inc., under NASA contract NAS5–26555. This work also is based in part on observations made with the NASA/ESA/CSA James Webb Space Telescope. The data were obtained from the Mikulski Archive for Space Telescopes at the Space Telescope Science Institute, which is operated by the Association of Universities for Research in Astronomy, Inc., under NASA contract NAS5-03127 for JWST.

\section*{Data Availability}
All retrieved data are available on a reasonable request to the corresponding author. The transmission spectra of WASP-12b and WASP-121b analysed in this study are obtained from the published work of \citet{2022ApJS..260....3C} and \citet{Saha_2025} respectively. 
 



\bibliographystyle{mnras}
\bibliography{My_ref} 








\bsp	
\label{lastpage}
\end{document}